\documentclass[12pt]{iopart}

\usepackage{iopams}  
\usepackage{braket}

\usepackage{tensor}
\usepackage{accents}
\usepackage{todonotes} 

\usepackage{xcolor}
\usepackage{subcaption}
\begin{document}

\title[About Jordan and Einstein frames: a study in inflationary magnetogenesis]{About Jordan and Einstein frames: a study in inflationary magnetogenesis}

\author[a,b,c]{Joel Vel\'asquez $^{1}$, H\'ector J. Hortua $^{2,3}$, L. Castañeda $^{1}$}

\address{$^1$Grupo de Gravitación y Cosmología, Observatorio Astronómico Nacional,\\
	Universidad Nacional de Colombia,	cra 45 \# 26-85,\ Ed.Uriel Gutierrez,\\
	Bogotá D.C., Colombia.\\
        $^2$Grupo Signos, Departamento de Matem\'aticas, Universidad el Bosque, Bogot\'a, Colombia.\\
        $^3$Maestr\'ia en Ciencia de Datos, Universidad Escuela Colombiana de Ingeniería Julio Garavito  Bogot\'a, Colombia.
 }

\ead{jjvelasquezc@unal.edu.co}
\vspace{10pt}
% \begin{indented}
% \item[]August 2017
% \end{indented}

\begin{abstract}
There has recently been considerable interest in the community to understand if the Einstein and Jordan frames are either physically equivalent to each other or if there exists  a preference frame where  interpretations of physical observables should be done. In this paper, we want to broaden the discussion about this  equivalence by making a detailed side-by-side comparison of the physical quantities in both frames in the context of cosmic magnetogenesis. We have computed the evolution of the vector potential in each frame along with  some observables such as  the spectral index, and the  magnetic field amplitude. We found that contrary to the Einstein frame, the electric and magnetic energy densities in Jordan Frame do not depend on any parameter associated with the scalar field. Furthermore, in the Einstein frame and assuming scale-invariant for the magnetic field, most of the total energy density contribution comes from the electric and magnetic densities. Finally, we show the ratio between magnetic field signals in both frames printed in the CMB. We expect that  the results presented contribute to the ongoing discussion on the relationship between these two frames.
\end{abstract}

%\maketitle

\section{Introduction}
\label{sec:intro}
One of the most exciting outstanding puzzles in modern Cosmology is the  origin of the accelerating expansion of the Universe~\cite{Riess,Perlmutter,Padmanabhan_2003,Carroll_2001}.  Modifications to Einstein's gravity have been  interesting candidates for explaining its origin without the cosmological constant. Representative modified gravity models that have been  studied include scalar-tensor theories (STTs) \cite{Brans_1961,Fujii:2003,Faraoni:2004,Quiros_2019, Carroll_2004, Chiba_2003, Nojiri_2003}, $f(R)$ gravity \cite{Carloni_2005, De_Felice_2010, Amendola_2007}, Gauss-Bonnet gravity \cite{Nojiri_2006, Amendola_2006, Fernandes_2022}, DGP (Dvali-Gabadadze-Porrati) model \cite{Dvali_2000}, brane-world gravity \cite{Maartens_2010} among others.    Depending on the coupling between the scalar field and the scalar curvature, STTs are formulated in two distinct frames, the Jordan Frame (JF) and the Einstein frame (EF). In the former, the scalar field is non-minimally coupled to gravity, while in the latter,  a minimal coupling is present. Both frames are related by  conformal transformations of the metric along with a scalar field redefinition. Moving  from JF to EF gets rid of   non-minimal coupling from the gravity sector in the action, and the Lagrangian of the redefined scalar restores its canonical form. Whereas, this transformation preserves  the non-minimal coupling  with the "new" scalar field in the matter sector. As a consequence, the matter energy-tensor momentum is no longer covariantly conserved implying that massive particles will not follow geodesics due to the appearance of an additional force in this frame. \\
 \textit{As a result, there is a perennial debate about whether one of them  provides a  physically privileged frame, or possibly  both represent the same theory, i.e., they are physically equivalents \cite{Faraoni_1999,Capozziello_1997,Flanagan_2004,Quiros_2012,Rashidi_2018, Rondeau_2017}. Despite  its mathematical equivalence has been accepted at least at the classical level without surface terms \cite{Postma_2014,Morris_2014,Nayem_2017}, the controversy  about its "true" equivalence is still present}.  \\
   In \cite{Nayem_2017}, the authors claim that both frames produce different symmetries that might not be translated back and forth using the transformation conformal, breaking the equivalence at least in the context of Noether symmetry. Nevertheless,  \cite{Nayem_2018,Nayem_2020} mentioned that the apparent non-equivalence is due to the fact that Noether theorem is not on-shell for constrained systems such as gravity. Also, as pointed out by \cite{Rinaldi_2018}, in quadratic and scale-invariant gravity the solution space of the Jordan frame cannot be entirely mapped into the solution space of the Einstein frame. Furthermore, it is possible to have acceleration in the Jordan frame, and when a conformal transformation moves  to the Einstein frame, the transformed metric can describe a decelerating Universe~\cite{Bahamonde_2017}.  
In \cite{Francfort_2019}, they present that gauge invariance does not guarantee frame invariance (e.g., the Bardeen potentials). Nevertheless, some authors claim the equivalence between both frames can be represented as a conformal transformation in a change of scale in the units of mass, time, and length \cite{Dicke_1962,Faraoni_2007,Postma_2014}. 
On the other hand, the authors in \cite{Racioppi_2022, Kuusk_2016, Karam_2017}  reported  that differences in the number of e-folds between the Jordan and Einstein frames can be quite significant, depending on the model of inflation. This difference might impact some inflation predictions such as an amplitude of the primordial curvature perturbations. Also, in the context of the Dark Energy Model \cite{NOZARI_2009}, the authors have shown that a non-minimally coupled scalar field in Jordan Frame resembles the dark energy component with the capability to realize phantom divide line crossing, although  its conformal transformation in the Einstein frame does not have this capability.
The aim of this paper is to analyze in a cosmological context  the equivalence between Jordan and Einstein frames, determine  whether one of the two is physically preferred, and illustrate their differences when  predictions or constraints are made.
To do so, we have chosen the scenario of magnetogenesis as the groundwork where nonminimal couplings and  conformal invariance breaking of the electromagnetic field are presented \cite{Subramanian_2016,Hort_a_2017,Caprini_2014, Bamba_2021, bamba2022helical}.
This paper is organized as follows: In  Sec.~\ref{CTsection}, we  briefly review  the conformal transformations between Jordan and Einstein frames. Sec.~\ref{u(1)_STTs}, describes the evolution of the potential vector using both frames and we evaluate power spectra. In sec.~\ref{Magnetogenesis}  we show different constraints on the magnetic field spectra in both frames while in Sec.~\ref{discussionFrames}  we discuss the CMB signal  left from those fields. Finally, we shall conclude with a summary in Sec.~\ref{conclusions}.
%======================================================================================================
\section{Conformal transformations between Jordan and Einstein frames}\label{CTsection}
In this section, we briefly show the standard procedure to demonstrate the equivalence between  Scalar-Tensor theories in the  Jordan and Einstein frames  \cite{Fujii:2003, De_Felice_2010, Faraoni:2004}.

\noindent Let us consider the  action for the nonminimally coupled scalar-tensor theories in the so-called Jordan frame\cite{Faraoni:2004, Nojiri_2017}
\begin{eqnarray}\label{eq:action_JF}
S_{J}=\int d^{4}x\sqrt{-g}Z(\phi,R),\\
Z=\frac{1}{2}f(\phi)R-\frac{1}{2}\omega(\phi)g^{cd}\nabla_{c}\phi\nabla_{d}\phi-V(\phi),
\end{eqnarray}
where the function $f(\phi)$ is the coupling function, $\omega(\phi)$ is a parameter and $V(\phi)$  is the potential of the scalar field. Performing a conformal transformation on the metric
\begin{equation}\label{eq:conf_transformation}
\accentset{\ast}{g}_{ab}=\Omega^{2}(x)g_{ab}, \quad \mbox{where} \quad\Omega^{2}=f(\phi),
\end{equation}
and defining a new scalar field $\chi$ \cite{Fujii:2003}
\begin{equation}\label{eq:chitophi}
\frac{d\chi}{d\phi}=\sqrt{\frac{3}{2}\left(\frac{f_{\phi}}{f}\right)^{2}+\frac{\omega}{f}},
\end{equation}
allows us to write the action in the Einstein frame \cite{Fujii:2003}
\begin{eqnarray}\label{eq:action_EF}
	S_{E}=\int d^{4}x\sqrt{-{}\accentset{\ast}g}Q(\chi,{}\accentset{\ast}{R}),\\
	Q=\frac{1}{2}{}\accentset{\ast}{R}-
	\frac{1}{2}{}\accentset{\ast}{g}^{cd}\nabla_{c}\chi
	\nabla_{d}\chi-U(\chi),
\end{eqnarray}
where ${}\accentset{\ast}{R}$ is the Ricci scalar corresponding to the metric ${}\accentset{\ast}{g}_{ab}$ and 
\begin{equation}
	U(\chi)=\frac{V(\phi(\chi))}{f(\phi(\chi))^{2}}.
\end{equation}
The  $f(R)$ gravity can be cast in the form of Scalar-Tensor theories considering the action without the kinetic term ($\omega(\phi)=0$) of the scalar field \cite{Velasquez_2020, Nojiri_2011}
\begin{equation}
	S_{f(R)}=\int d^{4}x\sqrt{-g}\left(f_{\phi}(R-\phi)+f(\phi)\right),
\end{equation}
and by taking the variation of the action with respect to the scalar field, we obtain
\begin{equation}
	f_{\phi\phi}(R-\phi)=0.
\end{equation}
Iff $f_{\phi\phi}\neq0$ then $\phi=R$, recovering the $f(R)$ action \cite{Bahamonde_2016}. Using this transformation,  the potential follows
\begin{equation}
V=\phi f_{\phi}-f(\phi)\longrightarrow V=Rf_{R}-f(R).
\end{equation}
We can get the Brans-Dicke (BD) theory, which is a particular case of the Scalar-Tensor theories via
\begin{equation}
f(\phi)=\phi,\quad\omega(\phi)=\frac{\omega_{BD}}{\phi},
\end{equation}
and plugging it into eq.(\ref{eq:chitophi}) becomes
\begin{equation}
\phi=e^{\sqrt{\frac{2}{3}}\chi},
\end{equation}
where we have used $\omega_{BD}=0$ because of the equivalence. In the literature, there is vast debate about whether the Jordan frame and Einstein frame are physically equivalent, i.e, if both frames are two distinct representations of the same theory or they do not provide any  physical equivalence \cite{Faraoni_1999,Faraoni:2004}.

%=====================================================================================================
\section{U(1) gauge field coupled with Scalar-Tensor Theories}\label{u(1)_STTs}
In what follows, we shall describe the magnetogenensis approach  in both Jordan and Einstein  Frames.   We will work on both frames independently in order to review the advantages and properties that each frame offers. 
\subsection{Magnetogenesis in Jordan Frame}
We consider a model with non-minimal coupling between Scalar-Tensor theories and the electromagnetic field in the Jordan frame
\begin{eqnarray}\label{eq:ActionintJ}
\fl \hspace{1cm}S_{int}^{J}=-\frac{1}{4}\int 
d^{4}x\sqrt{-g}Z(\phi,R)F_{ab}F^{ab}+\frac{\gamma_{g}}{4}\int 
d^{4}x\sqrt{-g}Z(\phi,R) F_{ab}\tilde{F}^{ab},
\end{eqnarray}
where $F_{ab}=\nabla_{a}A_{b}-\nabla_{b}A_{a}$ is the 
electromagnetic field-strength tensor. Here, $A_{a}$ is the $U(1)$ gauge field 
and $\tilde{F}^{ab}$ is the dual electromagnetic tensor. To obtain the equation of motion we vary the action with respect to $A_{b}$ 
\begin{eqnarray}
\frac{1}{\sqrt{-g}}\partial_{a}\left[\sqrt{-g}Z(\phi,R)\left( F^{
ab } -\frac{\gamma_{g}}{2}\epsilon^{abcd}F_{cd}\right)\right]=0,
\end{eqnarray}
where $\epsilon^{abcd}$ is the totally antisymmetric tensor defined as $\epsilon^{abcd}=\frac{\eta^{abcd}}{\sqrt{-g}}$. Here, $\eta^{abcd}$ is levi-cicita symbol. Working in the Coulomb gauge $A_{0}=0$, $\partial_{i}A^{i}=0$ the equation of motion is written as 
\begin{equation}
A''_{i}+\frac{Z'}{Z}A'_{i}-a^{2}(\tau)\partial^{j}\partial_{j}A_{i}+\frac
{Z'}{Z}\gamma\eta_{ijk}a^{2}(\tau)\partial^{j}A^{k}=0,
\end{equation}
where we have assumed the spatially flat Friedmann-Leamitre-Robertson-Walker (FLRW) spacetime 
\begin{equation}
ds^{2}=a^{2}(\tau)(-d\tau^{2}+d\mathbf{x}^{2}).
\end{equation}
Defining $\bar{A}_{i}=2\sqrt{Z}A_{i}$, the equation of motion   reads as
\begin{equation}\label{eq35}
\bar{A}''_{i}+\frac{1}{4}\left[\left(\frac{Z'}{Z}\right)^{2}-2\frac{Z''}{
Z}\right]\bar{A}_{i}-a^{2}(\tau)\partial^{j}\partial_{j}\bar{A}_{i}+\frac{Z'}{Z}
\gamma_{g}\eta_{ijk}a^{2}(\tau)\partial^{j}\bar{A}^{k}=0.
\end{equation}
Quantizing the electromagnetic field, we can expand the vector potential in the 
helicity basis in terms of creation and annihilation operators $\hat{b}^{\dagger}_{h}(k)$ and $\hat{b}_{h}(k)$ with the co-moving wave vector \cite{Subramanian_2010, Markkanen_2017, Sharma_2018},
\begin{equation}
\fl \hat{A}_{i}(\tau,\vec{x})=\int\frac{d^{3}k}{
	(2\pi)^{3/2}}\sum_{h=\pm}\Big[e_{i 
	h}(k)\hat{b}_{h}(k)A_{h}(\tau,\vec{x})e^{i\vec{k}\cdot 
	\vec{x}}
+e^{*}_{i 
h}(k)\hat{b}^{\dagger}_{h}(k)A_{h}^{*}(\tau,\vec{x})e^{-i\vec{k}\cdot 
	\vec{x}}\Big].
\end{equation}
Using the above expression along with  $\mathcal{A}=a(\tau)\bar{A}$, eq.~\ref{eq35} becomes 
\begin{equation}
\label{eq:evolution}
\mathcal{A}''_{h}+\left[k^{2}+\frac{Z'}{Z}\gamma_{g} h 
k+\frac{1}{4}\left(\frac{Z'}{Z}\right)^{2}	
-\frac{1}{2}\frac{Z''}{Z}\right]\mathcal{A}_{h}=0.
\end{equation}

 The evolution of this equation develops in three stages. At early times $k|\tau|\gg 1$ the term $k^{2}$ dominates over the last two (the mode is far inside the horizon). Later on, when $k|\tau|\ll 1$, the term proportional to $\gamma_{g}$ dominates, but only the modes $\gamma_{g}h > 0$ are amplified. Finally, as $\tau\rightarrow0$ the terms $\propto 1/\tau^{2}$ are amplified but the term $\gamma_{g}h<0$ is less amplified than the other case, for that reason we will neglect its effect \cite{Caprini_2014, Durrer_2022}. Now, before calculating the spectral densities of the electric and magnetic energy densities, we need to compute the contribution to the energy density of the electromagnetic field, to  achieve this, we will find the stress-energy tensor of the EM field which is obtained by varying the action equation (\ref{eq:ActionintJ}) with respect to the metric $g_{ab}$
\begin{eqnarray}
\fl T_{ab}=-\frac{2}{\sqrt{-g}}\frac{\delta 
S^{(JF)}}{\delta g^{ab}}=-\frac{1}{4}Z(\phi,R)g_{ab}F^{2}+Z(\phi,R)g^
{cd}F_{ac}F_{bd}\nonumber\\
\fl\qquad+\frac{1}{4}\left[f(\phi)F^{2}R_{ab}-g_{ab}
\Box\left(f(\phi)F^{2}\right)+\nabla_{a}\nabla_{b}\left(f(\phi)F^{2}
\right)\right]
-\frac
{1}{4}\omega(\phi)\nabla_{a}\phi\nabla_{b}\phi F^{2}\nonumber\\
\fl\qquad-\frac{\gamma}{4}\left[f(\phi)\tilde{F}^{2}R_{ab}-g_{ab}
\Box\left(f(\phi)\tilde{F}^{2}\right)+\nabla_{a}\nabla_{b}\left(f(\phi)\tilde{F}
^{2} \right)\right]+\frac
{\gamma}{4}\omega(\phi)\nabla_{a}\phi\nabla_{b}\phi \tilde{F}^{2},
\end{eqnarray}

where $F^{2}=F_{cd}F^{cd}$ and $\tilde{F}^{2}=F_{cd}\tilde{F}^{cd}$. Taking 
$a=b=0$, we have
\begin{eqnarray}
\fl T_{00}=\frac{1}{2}m_{1}g^{ij}A'_{i}A'_{j}
+\frac{1}{2}a^{2}m_{2}g^{ij}g^{kl}\partial_{j}A_{l}\left(\partial_{i}A_{k}
-\partial_{k}A_{i}\right)+\left(m_{3}a^{-2}g^{ij}A'_{i}A'_{j}\right)'\nonumber\\
\fl\qquad-\left(m_{3}g^{ij}g^{kl}\partial_{j}A_{l}\left(\partial_{i}A_{k}
-\partial_{k}A_{i}\right)\right)'+2\gamma_{g} 
m_{4}\epsilon^{ijk}A'_{i}
\partial_{j}A_{k}+2\gamma_{g}\left(m_{3}\epsilon^{ijk}A'_{i}
\partial_{j}A_{k}\right)'
\end{eqnarray}
where we have neglected the second-order spatial derivative of the quadratic quantity of electromagnetic fluctuations  \cite{Bamba_2008} and  defined the following quantities
\begin{eqnarray}
m_{1}\equiv\frac{1}{2}\left(f(\phi)R+3a^{-2}\omega(\phi)(\phi')^{2}
-2V(\phi)\right)\\
m_{2}\equiv\frac{1}{2}\left(f(\phi)R-a^{-2}\omega(\phi)(\phi')^{2}
-2V(\phi)\right)\\
m_{3}\equiv\frac{3}{2}\mathcal{H}f(\phi)\\
m_{4}\equiv\frac{1}{2}\omega(\phi)(\phi')^{2}.
\end{eqnarray}
Taking the expectation value for the stress-energy tensor in the vacuum state 
$\ket{0}$ (defined by the condition $b_{h}(k)\ket{0}=0$, for all $k$), we 
obtain the following
\begin{eqnarray}
 \fl-\Braket{0|\tensor{T}{^{0}_{0}^{(JF)}}|0}=\frac{m_{1}}{8\pi^{2}}\int_{0}^{
\infty} \frac{dk}{k}\frac{k^{3}}{a^{4}}
\left[\left|\left(\frac{\mathcal{A}_{+}(\tau,k)}{\sqrt{Z}}\right)'\right|^{2}
+\left|\left(\frac{\mathcal{A}_{-}(\tau,k)}{\sqrt{Z}}\right)'\right|^{2}\right]
\nonumber\\
\qquad+\frac{m_{2}}{8\pi^{2}}\int\frac{dk}{k} 
\frac{k^{5}}{a^{4}}
\left[\left|\frac{\mathcal{A}_{+}(\tau,k)}{\sqrt{Z}}\right|^{2}
+\left|\frac{\mathcal{A}_{-}(\tau,k)}{\sqrt{Z}}\right|^{2}\right]
\nonumber\\
\qquad+\frac{3
}{8\pi^{2}a^{2}}\frac{d}{d\tau}\int_{0}^{
\infty} \frac{dk}{k}\frac{k^{3}}{a^{4}}\mathcal{H}f(\phi)
\left[\left|\left(\frac{\mathcal{A}_{+}(\tau,k)}{\sqrt{Z}}\right)'\right|^{2}
+\left|\left(\frac{\mathcal{A}_{-}(\tau,k)}{\sqrt{Z}}\right)'\right|^{2}\right]
\nonumber\\
\qquad-\frac{3
}{8\pi^{2}a^{2}}\frac{d}{d\tau}\int_{0}^{
\infty}\frac{dk}{k} 
\frac{k^{5}}{a^{4}}\mathcal{H}f(\phi)\left[\left|\frac{\mathcal{A}_{+}(\tau,k)}{
\sqrt{ Z}}\right|^{2}
+\left|\frac{\mathcal{A}_{-}(\tau,k)}{\sqrt{Z}}\right|^{2}\right]\nonumber\\
\qquad+\frac{3
}{8\pi^{2}a^{2}}\frac{d}{d\tau}\int_{0}^{
\infty}\frac{dk}{k}\frac{k^4}{a^{4}}\mathcal{H}f(\phi)
\left(\left|\frac{\mathcal{A}_{+}(\tau,k)}{\sqrt{Z}}
\right|^{2}-\left|\frac{\mathcal{A}_{-}(\tau,k)}{\sqrt{Z}}
\right|^{2}\right)'
\end{eqnarray}

We can associate the first term of the above equation with the electric energy density stored at a given scale, the second term as magnetic energy density, and the rest as additional contributions to the total energy density.

\subsection{Magnetogenesis in Einstein Frame}
Let us follow the same procedure to calculate both the evolution equation of the potential vector and the energy density in the Einstein frame following the same procedure described in the previous section. The action in this frame is written as
\begin{equation}
S_{int}^{E}=-\frac{1}{4}\int 
d^{4}x\sqrt{-\accentset{\ast}{g}}Q(\chi,\accentset{\ast}{R})F_{ab}{}\accentset{\ast}{F}^{ab}
\nonumber\\
+\frac{1}{4}\int 
d^{4}x\sqrt{-\accentset{\ast}{g}}Q(\chi,\accentset{\ast}{R})\gamma_{g}F_{ab}{}\tilde{\accentset{
\ast}{
F}}^{ab}.
\end{equation}
Now, the equation of motion  for the  electromagnetic vector potential in the Coulomb gauge is given by     
\begin{equation}
\mathcal{A}''_{h}+\left[k^{2}+\frac{Q'}{Q}
\gamma\eta_{
ijk}k+\frac{1}{4}\left(\frac{Q'}{Q}\right)^{2}
-\frac{1}{2}\frac{Q''}{Q}\right]\mathcal{A}_{h}=0,
\end{equation}
where $\mathcal{A}=2a(\tau)\sqrt{Q}A_{i}$. The stress-tensor energy reads
\begin{eqnarray}
\hspace*{-1cm}\accentset{\ast}{T}_{ab}=-\frac{1}{4}Q(\chi,\accentset{\ast}{R})\accentset{\ast
} {g}_{ab}\accentset{\ast}{
F}^{2}+Q(\chi,\accentset{\ast}{R})\accentset{\ast}{g}^{cd}F_{ac}F_{bd}+\frac{1}{
4}\left(\accentset{\ast}{
F}^{2}\accentset{\ast}{
R}_{ab}-\accentset{\ast}{
g}_{ab}\accentset{\ast}{\Box}\accentset{\ast}{
F}^{2}+\accentset{\ast}{
\nabla}_{a}\accentset{\ast}{
\nabla}_{b}\accentset{\ast}{
F}^{2}\right)\nonumber\\
-\frac{\gamma}{
4}\left(\tilde{\accentset{\ast}{
F}}^{2}\accentset{\ast}{
R}_{ab}-\accentset{\ast}{
g}_{ab}\accentset{\ast}{\Box}\tilde{\accentset{\ast}{
F}}^{2}+\accentset{\ast}{
\nabla}_{a}\accentset{\ast}{
\nabla}_{b}\tilde{\accentset{\ast}{
F}}^{2}\right)-\frac{1}{4}\accentset{\ast}{\nabla}_{a}\chi\accentset{\ast}{
\nabla}_{b}\chi\accentset{\ast}{
F}^{2}+\frac{\gamma}{4}\accentset{\ast}{\nabla}_{a}\chi\accentset{\ast}{
\nabla}_{b}\chi\tilde{\accentset{\ast}{
F}}^{2},
\end{eqnarray}
where the time-time component is given by
\begin{eqnarray}
\hspace*{-1.5cm}\accentset{\ast}{T}_{00}=\frac{1}{2}\accentset{\ast}{m}_{1}\accentset{\ast}{g}^
{ij}A'_{i}A'_{j}+\frac{1}{2}\accentset{\ast}{m}_{2}\accentset{\ast}{a}^{2}
\accentset{\ast}{g}^{ij}\accentset{\ast}{g}^{kl}
\partial_{j}A_{l}\left(\partial_{i}A_{k}-\partial_{k}A_{i}\right) 
+\left(\accentset{\ast}{m}_{3}\accentset{\ast}{a}^{-2}\accentset{\ast}{g}^
{ij}A'_{i}A'_{j}\right)'\nonumber\\
\hspace*{-0.8cm}-\left(\accentset{\ast}{m}_{3}\accentset{\ast}{g}^{ij}\accentset{\ast}{g}^
{kl}\partial_{j}A_{l}\left(\partial_{i}A_{k}-\partial_{k}A_{i}
\right)\right)'+2\gamma\accentset{\ast}{m}_{4}\accentset{\ast}{
\epsilon}^{
ijk}
A'_{i}\partial_{j}A_{k}+2\gamma\left(\accentset{\ast}{m}_{3}\accentset{\ast}{
\epsilon }^{
ijk}
A'_{i}\partial_{j}A_{k}\right)',
\end{eqnarray}
and where we have defined the following functions
\begin{eqnarray}
\accentset{\ast}{m}_{1}\equiv\frac{1}{2}\left(\accentset{\ast}{R}
+3\accentset{\ast}{a}^{-2}\chi'^{2}-2U(\chi)\right)=Q+\frac{\chi'^{2}}{\accentset{\ast}{a}^{2}},\\
\accentset{\ast}{m}_{2}\equiv\frac{1}{2}\left(\accentset{\ast}{R}
-\accentset{\ast}{a}^{-2}\chi'^{2}
-2U(\chi)
\right)=Q-\frac{\chi'^{2}}{\accentset{\ast}{a}^{2}}.
\end{eqnarray}
Taking the expectation value for the stress-energy tensor in the vacuum state, we 
obtain the following terms
\begin{eqnarray} 
\fl\accentset{\ast}{\rho}_{E}=-\Braket{0|\tensor{\accentset{\ast}{T}}{^{0}_{0}^{(E)}}|0}=\frac{\accentset{
\ast}{m}_{1}}{8\pi^{2 }}\int_{0}^{
\infty} \frac{dk}{k}\frac{k^{3}}{\accentset{\ast}{a}^{4}}
\left[\left|\left(\frac{\accentset{\ast}{\mathcal{A}}_{+}(\accentset{\ast}{\tau},k)}{\sqrt{Q}}
\right)'\right|^ { 2 }
+\left|\left(\frac{\accentset{\ast}{\mathcal{A}}_{-}(\accentset{\ast}{\tau},k)}{\sqrt{Q}}
\right)'\right|^{2}\right],
\\
\fl\accentset{\ast}{\rho}_{B}=-\Braket{0|\accentset{\ast}{T}\indices{^0_0^{(B)}}|0}=\frac{\accentset{\ast}{m}
_{2}}{8\pi^{2}} \int\frac{dk}{k} 
\frac{k^{5}}{\accentset{\ast}{a}^{4}}
\left[\left|\frac{\accentset{\ast}{\mathcal{A}}_{+}(\accentset{\ast}{\tau},k)}{\sqrt{Q}}\right|^{2
}
+\left|\frac{\accentset{\ast}{\mathcal{A}}_{-}(\accentset{\ast}{\tau},k)}{\sqrt{Q}}\right|^{2}
\right],\\
\fl\Delta\accentset{\ast}{\rho}=\frac{3}{8\pi^{2}\accentset{\ast}{a}^{2}}\frac{d}{d\accentset{\ast}{\tau}}
\int\frac{dk}{k}\frac{k^{3}}{\accentset{\ast}{a}^{4}}\accentset{\ast}{\mathcal{H}}\left[\left|\left(\frac{\accentset{\ast}{\mathcal{A}}_{+}(\accentset{\ast}{\tau},k)}{\sqrt{Q}}
\right)'\right|^ { 2 }
+\left|\left(\frac{\accentset{\ast}{\mathcal{A}}_{-}(\accentset{\ast}{\tau},k)}{\sqrt{Q}}
\right)'\right|^{2}\right]\nonumber\\
\hspace*{-1.2cm}-\frac{3}{8\pi^{2}\accentset{\ast}{a}^{2}}\frac{d}{d\accentset{\ast}{\tau}}
\int\frac{dk}{k}\frac{k^{5}}{\accentset{\ast}{a}^{4}}\accentset{\ast}{\mathcal{H}}
\left[\left|\frac{\accentset{\ast}{\mathcal{A}}_{+}(\accentset{\ast}{\tau},k)}{\sqrt{Q}}\right|^{2
}
+\left|\frac{\accentset{\ast}{\mathcal{A}}_{-}(\accentset{\ast}{\tau},k)}{\sqrt{Q}}\right|^{2}
\right]\nonumber\\
\hspace*{-1.2cm}+\frac{3\gamma_{g}}{8\pi^{2}\accentset{\ast}{a}^{2}}\frac{d}{d\accentset{\ast}{\accentset{\ast}{\tau}}}
\int\frac{dk}{k}\frac{k^4}{\accentset{\ast}{a}^{4}}\accentset{\ast}{\mathcal{H}}
	\left(\left|\frac{\accentset{\ast}{\mathcal{A}}_{+}(\accentset{\ast}{\tau},k)}{\sqrt{Q}}
	\right|^{2}-\left|\frac{\accentset{\ast}{\mathcal{A}}_{-}(\accentset{\ast}{\tau},k)}{\sqrt{Q}}
	\right|^{2}\right)'\nonumber\\
\hspace*{-1.2cm}+\frac{\gamma_{g}}{8\pi^{2}\accentset{\ast}{a}^{2}}
\int\frac{dk}{k}\frac{k^4}{\accentset{\ast}{a}^{6}}\chi'^{2}
\left(\left|\frac{\accentset{\ast}{\mathcal{A}}_{+}(\accentset{\ast}{\tau},k)}{\sqrt{Q}}
\right|^{2}-\left|\frac{\accentset{\ast}{\mathcal{A}}_{-}(\accentset{\ast}{\tau},k)}{\sqrt{Q}}
\right|^{2}\right)'.
\end{eqnarray}
The total energy density can be calculated by adding up all the energy densities, i.e.,\newline $\accentset{\ast}{\rho}=\accentset{\ast}{\rho}_{E}+\accentset{\ast}{\rho}_{B}+\Delta\accentset{\ast}{\rho}$.
It is important to bear in mind  that the action in this frame has been taken as independent of the Jordan one. By taking a conformal transformation in Eq.(\ref{eq:ActionintJ}), we arrives at
\begin{equation}
\fl S_{int}^{E}=-\frac{1}{4}\int f^{2}
d^{4}x\sqrt{-\accentset{\ast}{g}}Q(\chi,\accentset{\ast}{R})F_{ab}{}\accentset{\ast}{F}^{ab}
+\frac{1}{4}\int f^{2}
d^{4}x\sqrt{-\accentset{\ast}{g}} Q(\chi,\accentset{\ast}{R})\gamma_{g}F_{ab}{}\tilde{\accentset{
\ast}{
F}}^{ab}.
\end{equation}
Notice the factor$f^2$ in the transformation which  reveals the conformal invariance breaking between both frames.
%======================================================================================================
\section{Magnetogenesis on Power-law inflation}\label{Magnetogenesis}
In the previous section, we obtained the spectral densities of the electric and magnetic densities of Jordan's and Einstein's frames. In this section, we want to explore more about the above results using a specific model for these frames.
\subsection{Model in Jordan Frame}

By using the magnetogenesis procedure, which assumes that the coupling functions evolve by a power law, we assume the evolution of the coupling $Z$ in this manner because, in this context, we are interested in the asymptotic  solutions
\begin{equation}\label{eq:evolveZmodel}
 Z=Z_{0}\left(\frac{\tau}{\tau_{0}}\right)^{-\gamma}.
\end{equation}

\begin{figure}[h]
	\centering % \begin{center}/\end{center} takes some additional vertical space
	\includegraphics[scale=0.5]{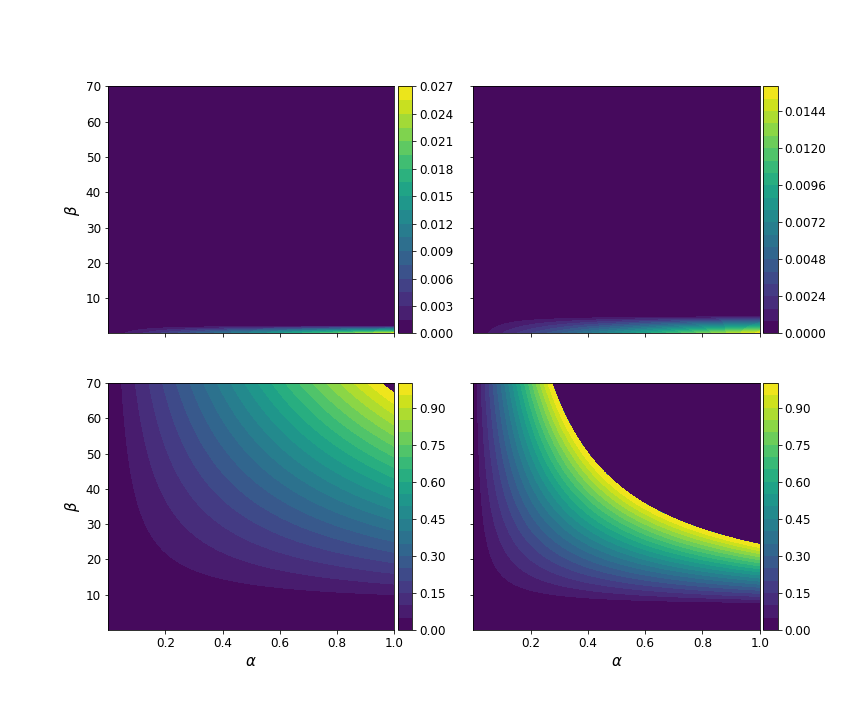}
	% "\includegraphics" is very powerful; the graphicx package is already loaded
	\caption{\label{fig:constraintallxi} Contour plots  of  the forbidden  regions (in purple) for $\Delta\rho$ for four different values of $\xi$. Left and right upper plots display the region for $\xi=-2$ and  $\xi=0$ respectively.  $\xi=1$ and $\xi=2$ are described in  the left and right bottom plots respectively.}
\end{figure}

The vector potential behaves
\begin{equation} 
\mathcal{A}_{h}''+\left(k^{2}-\frac{2\xi\gamma_{g}hk}{\tau}-\frac{\xi(\xi+1)}{
\tau^{2}}\right)\mathcal{A}_{h}=0,
\end{equation}
where $\xi=2\gamma$. The solution to this equation is given by \cite{olver10}
\begin{equation}
\mathcal{A}_{h}=C_{1}W_{\kappa,\mu}(z)+C_{2}W_{-\kappa,\mu}(z),
\end{equation}
being $W_{\kappa,\mu}(z)$ the Whittaker functions.  The asymptotic representations of these functions
are \cite{olver10}
\begin{eqnarray}
W_{\kappa,\mu}(z)=\left\{
\begin{array}{ll}
\frac{\Gamma(2\mu)}{\Gamma(\frac{1}{2}+\mu-\kappa)}z^{\frac{1}{2}-\mu}+\frac{
\Gamma(-2\mu)}{\Gamma(\frac{1}{2}-\mu-\kappa)}z^{\frac{1}{2}+\mu}, z\rightarrow 
0\\
e^{-\frac{1}{2}z}z^{\kappa}, z\rightarrow \infty
\end{array}
\right.
\end{eqnarray}
In order to determine the coefficients $C_{1}$ and $C_{2}$, we have to match the 
solution with the Bunch-Davies vacuum
\begin{equation}\label{eq:Bunch_Davies_V}
\mathcal{A}\rightarrow\frac{1}{\sqrt{2k}}e^{-ik\tau},\quad
 \mbox{for} \quad-k\tau\rightarrow\infty.
\end{equation}
As a result, we see that $C_{2}=0$ and $C_{1}$ becomes
\begin{eqnarray}
C_{1}=\frac{1}{\sqrt{2k}}e^{\xi h \gamma\pi/2}.
\end{eqnarray}
At the end of the inflation, all the modes outside the horizon will be 
given by
\begin{equation}
\mathcal{A}_{h}=\frac{e^{\xi h 
\gamma\pi/2}}{\sqrt{2k}}\left[\underbrace{\frac{(-2i)^{-\xi}\Gamma(2\xi+1)
}{\Gamma(\xi+1-ih\gamma\xi)}}_{C_{3}}(-k\tau)^{-\xi}+\underbrace{\frac{
(-2i)^{\xi+1}\Gamma(-2\xi-1)}{\Gamma(-\xi-ih\gamma\xi)}}_{C_{4}}
(-k\tau)^{\xi+1}\right].
\end{equation}
Assuming maximal helicity
$|A_{+}|=|A|$ and $|A_{-}|=0$,  the expectations value become
\begin{eqnarray}
\fl-\Braket{0|\tensor{T}{^{0}_{0}^{(JF)}}|0}=\frac{H^{4}}{16\pi^{2}}e^{
\pi\xi\gamma_{g}} \int_
{0}^{\infty}\frac{dk}{k}
\left|C_{4}\right|^{2}\left(\frac{k}{aH}\right)^{2\xi+4}
(2\xi+1)^{2}\nonumber\\
\hspace*{0.65cm}+\frac{H^{4}}{16\pi^{2}}e^{
\pi\xi\gamma_{g}}\int_
{0}^{\infty}\frac{dk}{k}
\left[\left|C_{3}\right|^{2}\left(\frac{k}{aH}\right)^{-2\xi+4}+\left|C_{4}
\right|^{2}\left(\frac{k}{aH}\right)^{2\xi+6}\right]\nonumber\\
\hspace*{0.65cm}+\frac{3H^{6}}{16\pi^{2}}e^{
\pi\xi\gamma_{g}}\beta_{0} \int_
{0}^{\infty}\frac{dk}{k}
\left|C_{4}\right|^{2}\left(\frac{k}{aH}\right)^{2\xi+4}
(2\xi+1)^{2}\left(\beta-(2\xi+3)\right)\tau^{-\beta}\nonumber\\
\hspace*{0.65cm}-\frac{3H^{6}}{16\pi^{2}}e^{
\pi\xi\gamma_{g}}\beta_{0}\int_
{0}^{\infty}\frac{dk}{k}
\Bigg[\left|C_{3}\right|^{2}\left(\frac{k}{aH}\right)^{-2\xi+4}
\left(\beta-(-2\xi+3)\right)\nonumber\\
\hspace{4cm}+\left|C_{4}
\right|^{2}\left(\frac{k}{aH}\right)^{2\xi+6}\left(\beta-(2\xi+5)\right)\Bigg]
\tau^{-\beta} \nonumber\\
\hspace*{0.65cm}+\frac{3H^{6}}{16\pi^{2}}e^{
\pi\xi\gamma_{g}}\beta_{0} \int_
{0}^{\infty}\frac{dk}{k}
\left|C_{4}\right|^{2}\left(\frac{k}{aH}\right)^{2\xi+5}
2(2\xi+1)\left(2\xi+4-\beta\right)\tau^{-\beta},
\end{eqnarray}
where we have assumed that the fraction $\frac{f(\phi)}{Z}=\beta_{0}\tau^{-\beta}$. The invariance scale in the magnetic field is given by  $\xi=2,-3$, but to the value $\xi=-3$ the electric field diverges as  $\left(\frac{k}{aH}\right)^{-2}$ in the super-horizon limit. To avoid an excessive production of electromagnetic energy, we calculate 
the energy stored in the electromagnetic field at the end of the inflation $\tau_{f}$
\begin{eqnarray}\label{eq:densityJF_Example}
\hspace*{-0.8cm}\rho^{(JF)}_{EM}=\frac{H^{4}}{16\pi^{2}}e^{
\pi\xi\gamma_{g}}\frac{(2\xi+1)^{2}}{2\xi+4}\left|C_{4}\right|^{2}
\left(1-e^{-(2\xi+4)N}\right)\nonumber\\
\hspace*{0.5cm}+\frac{H^{4}}{16\pi^{2}}e^{
\pi\xi\gamma_{g}}
\left[\frac{\left|C_{3}\right|^{2}}{-2\xi+4}\left(1-e^{
-(-2\xi+4)N}\right)+\frac{\left|C_ {4}
\right|^{2}}{2\xi+6}\left(1-e^{-(2\xi+6)N}\right)\right]\nonumber\\
\hspace*{0.5cm}+\frac{3H^{6}}{16\pi^{2}}e^{
\pi\xi\gamma_{g}}\beta_{0}\frac{(2\xi+1)^{2}}{2\xi+4}
\left(\beta-(2\xi+3)\right)\tau_{f}^{-\beta}
\left|C_{4}\right|^{2}\left(1-e^{-(2\xi+4)N}\right)
\nonumber\\
\hspace*{0.5cm}-\frac{3H^{6}}{16\pi^{2}}e^{
\pi\xi\gamma_{g}}\beta_{0}
\Bigg[\left|C_{3}\right|^{2}\frac{\beta-(-2\xi+3)}{-2\xi+4}
\left(1-e^{-(-2\xi+4)N}\right)
\nonumber\\
\hspace{4cm}+\left|C_{4}
\right|^{2}\frac{\beta-(2\xi+5)}{
2\xi+6}\left(1-e^{-(2\xi+6)N}\right) \Bigg ]
\tau_{f}^{-\beta} \nonumber\\
\hspace*{0.5cm}+\frac{3H^{6}}{16\pi^{2}}e^{
\pi\xi\gamma_{g}}\beta_{0}\left|C_{4}\right|^{2}
\frac{2(2\xi+1)\left(2\xi+4-\beta\right)}{2\xi+5}\left(1-e^{
-(2\xi+5)N}\right)\tau_{f}^{-\beta},
\end{eqnarray}
where $N$ is the number of e-folds and it is defined by $N\equiv\ln\frac{a_{f}}{a_{i}}$. \newline
 Using $\left.\frac{f(\phi)}{Z}\right|_{\tau = \tau_{f}}=\beta_{0}\tau_{f}^{-\beta}=\alpha H^{-2}$, being $\alpha$ a parameter that runs to $0-1$ to not spoil inflation energy.
 Figure \ref{fig:constraintallxi} displays the allowed area for $\Delta\rho$ constrained by the $\alpha$ and $\beta$ values.  $\alpha$ values run to $0-1$, while $\beta$ goes to $0-70$.  We can observe in the upper left plot for $\xi=-2$, that $\Delta\rho$  permits only  small $\beta$ values ($\beta\sim2$).

In contrast, for $\xi>0$ the allowed region becomes higher yielding a broad range for these parameters. 

\begin{figure}[ht]
	\centering % \begin{center}/\end{center} takes some additional vertical space
	\includegraphics[width=\textwidth]{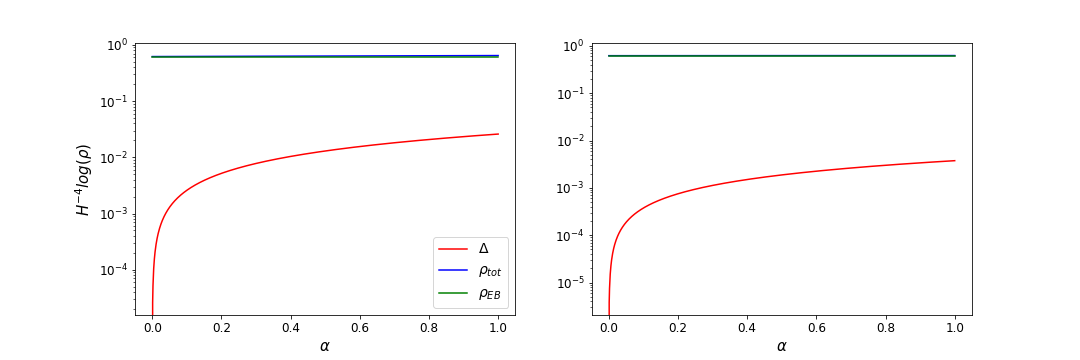}
	% "\includegraphics" is very powerful; the graphicx package is already loaded
	\caption{\label{fig:magnetic_density_xineg2} Plots for $\Delta\rho$ (red line), $\rho_{EB}=\rho_{E}+\rho_{B}$ (green line) and $\rho_{tot}$ (blue line) using $\beta=0.01$ (left plot) and  $\beta=1.9$ (right plot) for $\xi=-2$.}
\end{figure}

\begin{figure}[h]
	\centering % \begin{center}/\end{center} takes some additional vertical space
	\includegraphics[width=\textwidth]{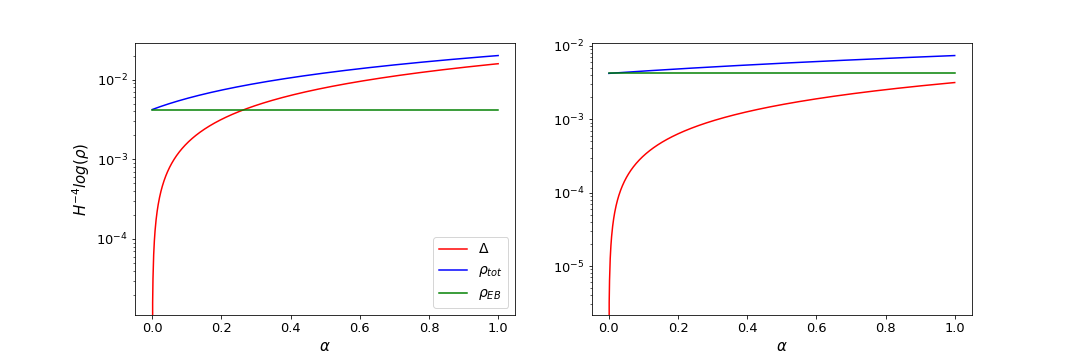}
	% "\includegraphics" is very powerful; the graphicx package is already loaded
	\caption{\label{fig:magnetic_density_xi0}  Plots for $\Delta\rho$ (red line), $\rho_{EB}=\rho_{E}+\rho_{B}$ (green line) and $\rho_{tot}$ (blue line) using $\beta=0.01$ (left plot) and  $\beta=4$ (right plot) for $\xi=0$.}
\end{figure}

\begin{figure}[h]
	\centering % \begin{center}/\end{center} takes some additional vertical space
	\includegraphics[width=\textwidth]{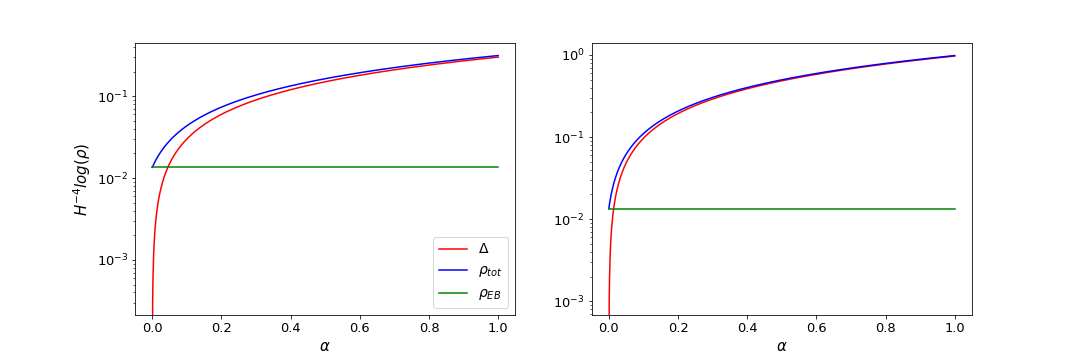}
	% "\includegraphics" is very powerful; the graphicx package is already loaded
	\caption{\label{fig:magnetic_density_xi1} Plot for $\Delta\rho$ (red line), $\rho_{EB}=\rho_{E}+\rho_{B}$ (green line) and $\rho_{tot}$ (blue line) using $\beta=25$ (left plot) and  $\beta=65$ (right plot) for $\xi=1$.}
\end{figure}

\begin{figure}[h]
	\centering % \begin{center}/\end{center} takes some additional vertical space
	\includegraphics[width=\textwidth]{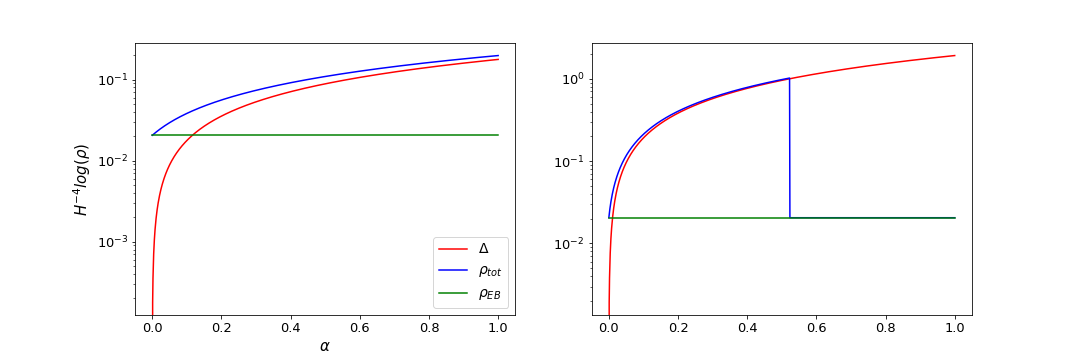}
	% "\includegraphics" is very powerful; the graphicx package is already loaded
	\caption{\label{fig:magnetic_density_xi2} Plot for $\Delta\rho$ (red line), $\rho_{EB}=\rho_{E}+\rho_{B}$ (green line) and $\rho_{tot}$ (blue line) using $\beta=10$ (left plot) and  $\beta=40$ (right plot) for $\xi=2$.}
\end{figure}

%For $\Delta\rho$ only $\beta$ small values are permitted, the contribution for $\rho_{tot}$ values are due to $\rho_{E}+\rho_{B}$
Figure \ref{fig:magnetic_density_xineg2} shows the behavior of $\Delta\rho$ and $\rho_{tot}$ for two $\beta$ values taken from the previous analysis using $\xi=-2$. The remaining energy density represented by the  red line  increases with $\alpha$ although its   contribution is negligible with respect to the  magnetic and electric densities.
\newpage
Figures \ref{fig:magnetic_density_xi0} and \ref{fig:magnetic_density_xi1}, illustrate the behavior of the electromagnetic field and $\Delta \rho$ for $\xi=0,1$ respectively. Notice how the combination for $\alpha$ and $\beta$ determines the larger contribution for either $\rho_{EB}$ or $\Delta\rho$. Finally, we can also approximate the forbidden  limit at which the total density equals the inflation energy as   it is shown in \ref{fig:magnetic_density_xi2}.
\newline

\newpage

Assuming that  the power spectra scale as a power law, we can write the  magnetic spectral index as
\begin{eqnarray}\label{eq:nBJF}
	2(2\pi)^{3}P_{B}=k^{2}<|A|^{2}>\nonumber\\
%	&=\frac{e^{\xi\gamma_{g}\pi}}{8\pi^{2}a^{2}Z}<\left[|C_{3}|^{2}(-k\tau)^{-2\xi+4}+|C_{4}|^{2}(-k\tau)^{2\xi+6}\right]>\nonumber\\
	\Rightarrow k^{n_{B}}\propto P_{k}\rightarrow n_{B}=-2\xi+1 \quad\mbox{for}-k\tau<<1.
\end{eqnarray}
where we have used the fact that the magnetic field strength as
\begin{equation}
\fl \hspace{1cm}B(k)=kA_{+}(k)=\frac{k\mathcal{A}_{+}(k)}{2a\sqrt{Z}}=\frac{e^{\pi\gamma_{g}}\sqrt{k}}{2\sqrt{2}a\sqrt{Z}}\left(C_{3}\left(-k\tau\right)^{-\xi}+C_{4}\left(-k\tau\right)^{\xi+1}\right).
\end{equation}
On the other hand, we can compute the strength of the magnetic field to the present day assuming that it is  scale-invariant and also that the universe is  instantaneously shifted from inflation to radiation domination \cite{Subramanian_2016}. Hence, the temperature at the end of inflation is
\begin{eqnarray}
\fl T_{f}=\left(\frac{90}{8\pi^{3}}\right)^{1/4}\frac{H^ {1/2}M_{p}^{1/2}}{T_{0}}\frac{g_{f}^{1/12}}{g_{0}^{1/3}}=\left(\frac{90}{8\pi^{3}}\right)^{1/4}\frac{10^{-5/2}M_{p}}{T_{0}}\frac{100^{1/12}}{2.64^{1/3}}\left(\frac{H}{10^ {-5}M_{p}}\right)^{1/2}\nonumber\\
	\hspace*{0.60cm}=0.0026\frac{M_{p}}{T_{0}}\left(\frac{H}{10^ {-5}M_{p}}\right)^{1/2}.
\end{eqnarray}
Since the magnetic density decreases with the expansion as $a^{-4}$, the value of the magnetic field for the actual epoch becomes
\begin{equation}\label{eq:B_actualepoch_JF}
	\rho_{B_{0}}=\rho_{B}\left(\frac{a_{f}}{a_{0}}\right)^4\rightarrow B_{0}=0.63\times10^{-10}G\left(\frac{H}{10^ {-5}M_{p}}\right)
\end{equation}
where have been used the entropy conservation 
\begin{equation}
	\frac{a_{0}}{a_{f}}=\left(\frac{g_{f}}{g_{0}}\right)^{1/3}\frac{T_{f}}{T_{0}},
\end{equation}
where $g_{f}\sim100$  and $g_{0}\sim2.64$ \cite{Subramanian_2016}.

Finally, the helicity can be found using the following equation
\begin{eqnarray}
\fl	\mathcal{H}=\int A\cdot B d^{3}x\Rightarrow\mathcal{H}=\frac{1}{(2\pi)^{3}}\int |A_{k}|^{2}k\ d^{3}k=\frac{1}{2\pi^{2}}\int k^{3}|A_{k}|^{2}dk\nonumber\\
\hspace*{1.6cm}=\frac{1}{8\pi^{2}a^{2}Z}\int k^{3}|\mathcal{A}_{k}|^{2}dk\\
\fl\mathcal{H}=\frac{e^{\xi\gamma_{g}\pi}}{16\pi^{2}a^{2}Z}\left[\frac{|C_{3}|^{2}}{-2\xi+3}(-k\tau)^{-2\xi}k^{3}+\frac{|C_{4}|^{2}}{2\xi+5}(-k\tau)^{2\xi+2}k^{3}\right]
\end{eqnarray}
being the kinetic helicity spectral index
\begin{equation}
	n_{H}=-2\xi+2.
\end{equation}

\subsection{Magnetogenesis  view from the  Einstein Frame}
Let us start assuming  a power-law for the $Q-$coupling
\begin{equation}
	Q=Q_{0}\left(\frac{\accentset{\ast}{\tau}}{\accentset{\ast}{\tau}_{0}}\right)^{-\eta}.
\end{equation}
We can see that $Q$ evolves the same way as $Z$ in Jordan Frame does in eq \ref{eq:evolveZmodel} in order to find asymptotic solutions and compare them with those found previously. Bear in mind this chosen ansatz will be used to analyze the magnetogenesis observables similar to the development made in the Jordan frame.\\

\begin{figure}[ht]
	\centering % \begin{center}/\end{center} takes some additional vertical space
	\includegraphics[width=\textwidth]{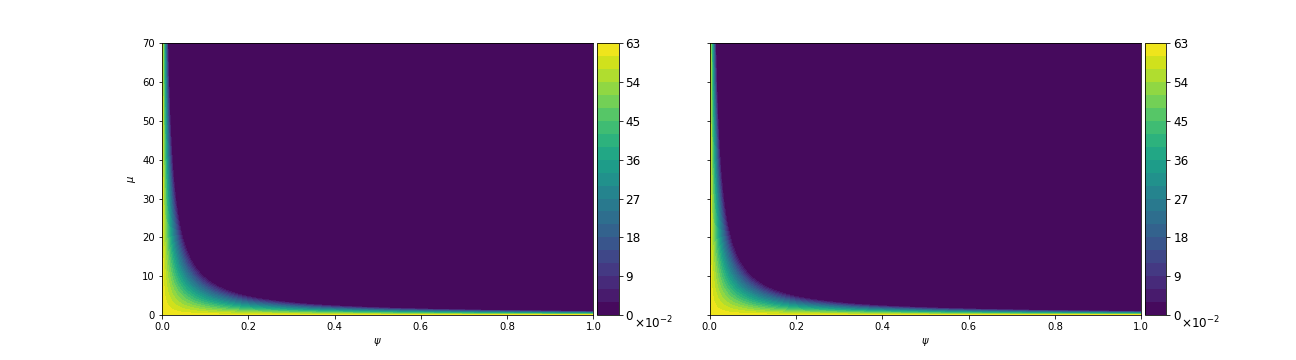}
	% "\includegraphics" is very powerful; the graphicx package is already loaded
	\caption{\label{fig:constraint_1EF} Contour plots display the forbidden regions for $\rho_{B}$ (left plot) and $\rho_{tot}$ (right plot) taking $\delta=-2$. $\psi$ runs from 0 to 1, $\mu$ ranges $0-70$ and $\nu=0.5$. Here, the amplitude scale is $\times10^{-2}$, and the forbidden values are shown in purple. We can see the greatest contribution comes from $\rho_{E}$.}
\end{figure}

\begin{figure}
	\centering % \begin{center}/\end{center} takes some additional vertical space
	\includegraphics[width=\textwidth]{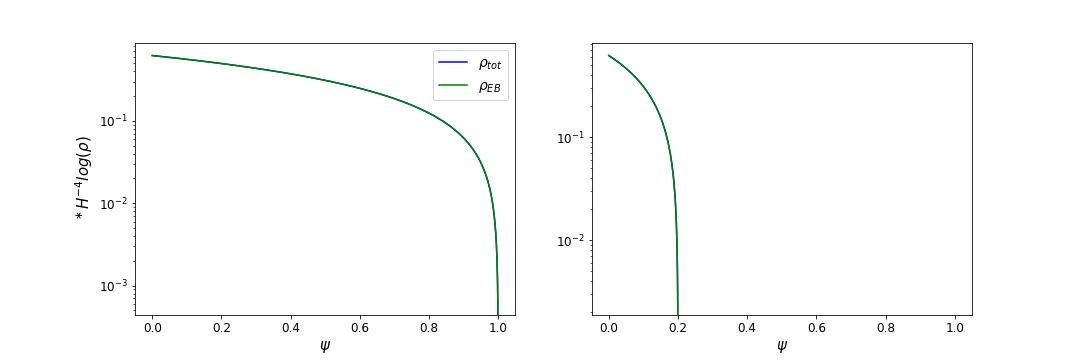}
	% "\includegraphics" is very powerful; the graphicx package is already loaded
	\caption{\label{fig:Energy_1EF} Plots for $\rho_{EB}=\rho_{E}+\rho_{B}$ (green line) and $\rho_{tot}$ (blue line) taking $\mu=1$ (left plot) and  $\mu=5$ (right plot). The remaining values are $\delta=-2$, $\nu=0.5$, and $\psi$ goes from $0$ to $1$.}
\end{figure}

\begin{figure}[!htb]
	\centering % \begin{center}/\end{center} takes some additional vertical space
	\includegraphics[width=\textwidth]{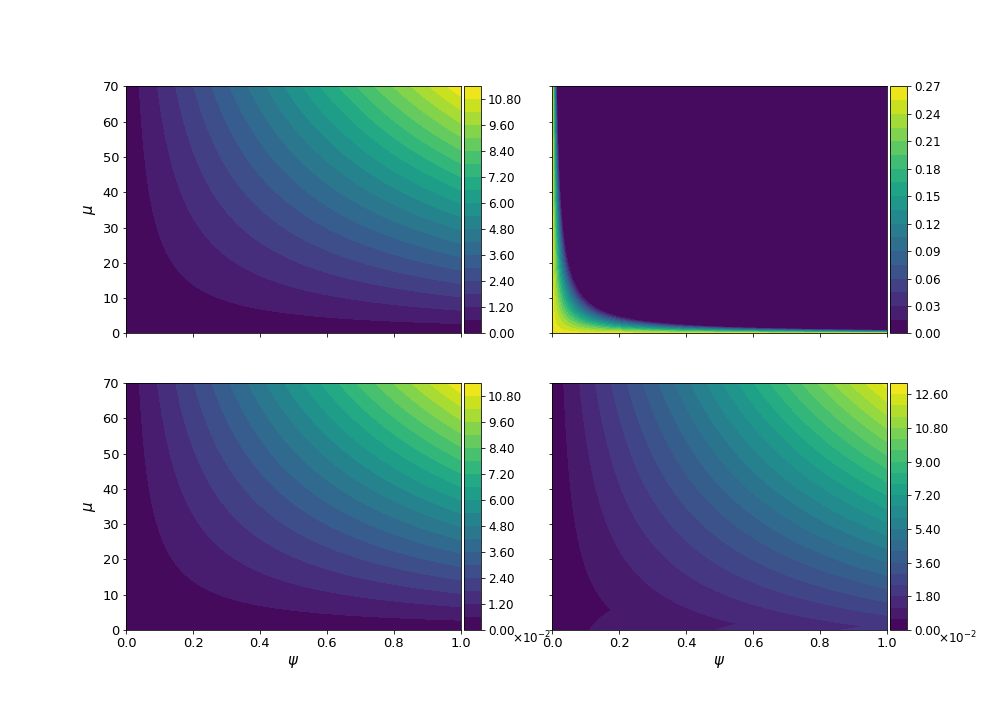}
	% "\includegraphics" is very powerful; the graphicx package is already loaded
	\caption{\label{fig:constraint_2EF} Contour plot for forbidden regions for $\rho_{E}$ (left upper plot), $\rho_{B}$ (right upper plot), $\rho_{EB}=\rho_{E}+\rho_{B}$ (left bottom plot) and $\rho_{tot}$ (right bottom plot) taking $\delta=0$. $\psi$ goes from 0 to 1, $\mu$ ranges $0-70$, and $\nu=0.5$. Here, the scale is $\times10^{-2}$, and the forbidden values are shown in purple.}
\end{figure}

\begin{figure}[!htb]
	\centering % \begin{center}/\end{center} takes some additional vertical space
	\includegraphics[width=\textwidth]{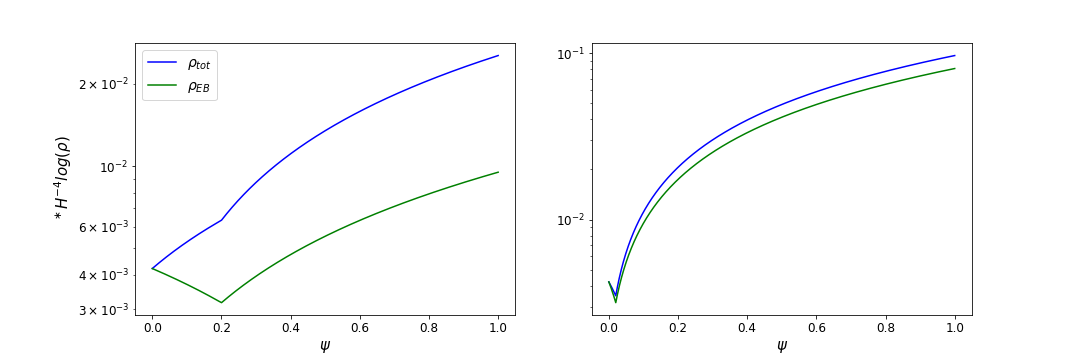}
	% "\includegraphics" is very powerful; the graphicx package is already loaded
	\caption{\label{fig:Energy_2EF}  Plots for $\rho_{EB}$ (green line), $\rho_{tot}$ (blue line) taking $\mu=5$ (left plot), and  $\mu=50$ (right plot) for $\delta=0$,  $\nu=0.5$, and $\psi$ ranges $0-1$.}
\end{figure}

\begin{figure}[!htb]
	\centering % \begin{center}/\end{center} takes some additional vertical space
	\includegraphics[width=\textwidth]{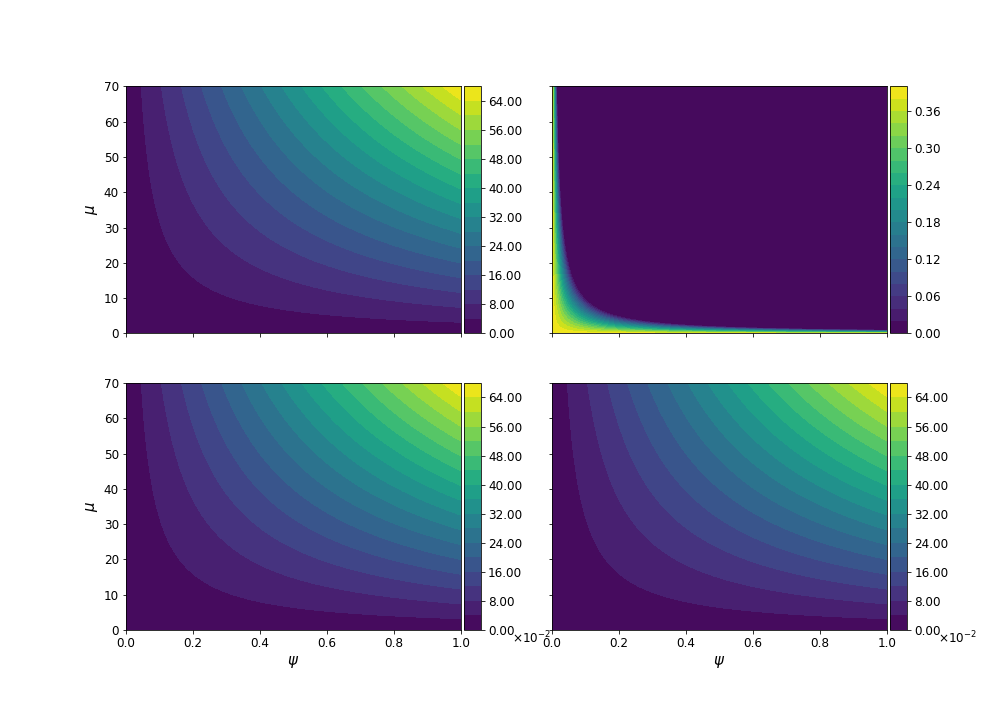}
	% "\includegraphics" is very powerful; the graphicx package is already loaded
	\caption{\label{fig:constraint_3EF} Contour plot with the permitted values and forbidden regions for $\rho_{E}$ (left upper plot), $\rho_{B}$ (right upper plot), $\rho_{EB}$ (left bottom plot) and $\rho_{tot}$ (right bottom plot) with $\delta=1$. $\psi$ runs from 0 to 1, $\mu$ ranges $0-70$. Here, the scale of the plots $\times10^{-2}$, and the forbidden values are shown in purple.}
\end{figure}

\begin{figure}[!htb]
\centering 
\includegraphics[width=\textwidth]{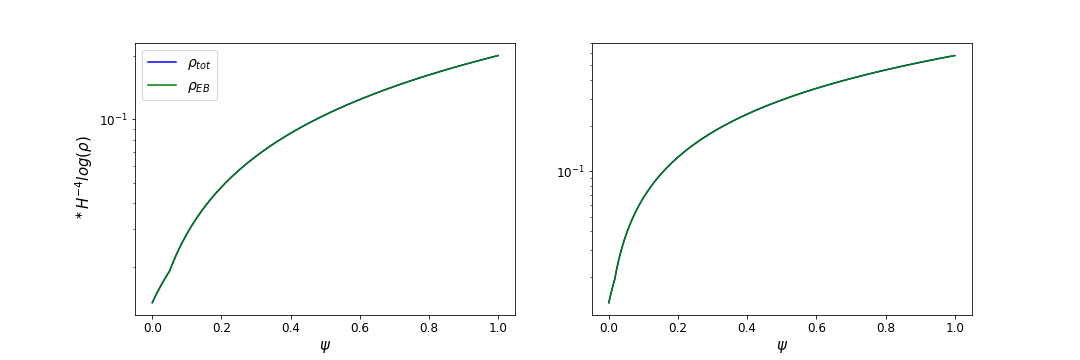}
\caption{\label{fig:Energy_3EF}  plot for $\rho_{EB}$ (green line) and $\rho_{tot}$ (blue line) taking $\mu=20$ (left plot) and  $\mu=60$ (right plot) for $\delta=1$ and choosing $\nu=0.5$, $\psi$ goes to $0-1$. }
\end{figure}

The evolution equation for the vector potential in this frame becomes
\begin{equation} 
	\mathcal{\accentset{\ast}{A}}_{h}''+\left(k^{2}-\frac{2\delta\gamma_{g}hk}{\accentset{\ast}{\tau}}-\frac{\delta(\delta+1)}{
		\accentset{\ast}{\tau}^{2}}\right)\mathcal{\accentset{\ast}{A}}_{h}=0,
\end{equation}
where $\delta=2\eta$.  After  following a straightforward procedure similar to the one used in the Jordan frame of the previous section,  we arrive at
\begin{eqnarray} 
\fl-\Braket{0|\tensor{\accentset{\ast}{T}}{^{0}_{0}^{(EF)}}|0}=\frac{\accentset{
\ast}{H}^{4}}{16\pi^ {2}}e^{
\pi\delta\gamma_{g}}\left(1+\frac{\accentset{
\ast}{a}^{-2}x'^{2}}{Q}\right) \int_
{0}^{\infty}\frac{dk}{k}
\left|C_{4}\right|^{2}\left(\frac{k}{\accentset{
\ast}{a}\accentset{
\ast}{H}}\right)^{2\delta+4}
(2\delta+1)^{2}\nonumber\\
+\frac{\accentset{
\ast}{H}^{4}}{16\pi^{2}}e^
{
\pi\delta\gamma_{g}}\left(1-\frac{\accentset{
\ast}{a}^{-2}x'^{2}}{Q}\right)\int_
{0}^{\infty}\frac{dk}{k}
\Big[\left|C_{3}\right|^{2}\left(\frac{k}{\accentset{
\ast}{a}\accentset{
\ast}{H}}\right)^{-2\delta+4}
\nonumber\\
\hspace{6cm}+\left|C_{4}\right|^{2}\left(\frac{k}{\accentset{
\ast}{a}\accentset{
\ast}{H}}\right)^{2\delta+6}\Big]\nonumber\\
-\frac{3\accentset{
\ast}{H}^{6}}{16\pi^{2}}e^{
\pi\delta\gamma_{g}} \int_
{0}^{\infty}\frac{dk}{k}
\frac{\left|C_{4}\right|^{2}}{Q}\left(\frac{k}{\accentset{
\ast}{a}\accentset{
\ast}{H}}\right)^{2\delta+4}
(2\delta+1)^{2}(4\delta+3)\nonumber\\
+\frac{3\accentset{
\ast}{H}^{6}}{16\pi^{2}}e^{
\pi\delta\gamma_{g}}\int_
{0}^{\infty}\frac{dk}{k}
\Bigg[3\frac{\left|C_{3}\right|^{2}}{Q}\left(\frac{k}{\accentset{
\ast}{a}\accentset{
\ast}{H}}\right)^{-2\delta+4}
+\frac{\left|C_{4}\right|^{2}}{Q}\left(\frac{k}{\accentset{
\ast}{a}\accentset{
\ast}{H}}\right)^{2\delta+6}
\left(4\delta+5\right)\Bigg] \nonumber\\
+\frac{12\gamma_{g}\accentset{
\ast}{H}^{6}}{16\pi^{2}}e^{
\pi\delta\gamma_{g}} \int_
{0}^{\infty}\frac{dk}{k}
\frac{\left|C_{4}\right|^{2}}{Q}\left(\frac{k}{\accentset{
\ast}{a}\accentset{
\ast}{H}}\right)^{2\delta+5}
2(2\delta+1)(\delta+1) \nonumber\\
+\frac{\gamma_{g}\accentset{
\ast}{H}^{6}}{16\pi^{2}}e^{
\pi\delta\gamma_{g}} \int_
{0}^{\infty}\frac{dk}{k}
\frac{\chi'^{2}}{Q}\left|C_{4}\right|^{2}\left(-\frac{1}{\accentset{
\ast}{a}\accentset{
\ast}{H}}\right)\left(\frac{k}
{\accentset{
\ast}{a}\accentset{
\ast}{H}} \right)^ { 2\delta+5 }
2(2\delta+1).
\end{eqnarray}
And where  the energy density at the end of inflation in this frame reads
\begin{eqnarray}\label{eq:densityEF_Example}
\fl\accentset{\ast}{\rho}_{EM}=\frac{H^{4}}{16\pi^{2}}e^{
\pi\delta\gamma_{g}}\frac{(2\delta+1)^{2}}{2\delta+4}\left|C_{4}\right|^{2}
\left(1+\psi\mu\right)\left(1-e^{-(2\delta+4)N}\right)\nonumber\\
\hspace*{-1.25cm}+\frac{H^{4}}{16\pi^{2}}e^{
\pi\delta\gamma_{g}}\left(1-\psi\mu\right)
\left[\frac{\left|C_{3}\right|^{2}}{-2\delta+4}\left(1-e^{
-(-2\delta+4)N}\right)+\frac{\left|C_ {4}
\right|^{2}}{2\delta+6}\left(1-e^{-(2\delta+6)N}
\right)\right] \nonumber\\
\hspace*{-1.25cm}-\frac{3H^{4}}{16\pi^{2}}e^{
\pi\delta\gamma_{g}}\psi\frac{(2\delta+1)^{2}}{2\delta+4
}
\left(4\delta+3\right)
\left|C_{4}\right|^{2}\left(1-e^{-(2\delta+4)N}\right)
\nonumber\\
\hspace*{-1.25cm}+\frac{3H^{4}}{16\pi^{2}}e^{
\pi\delta\gamma_{g}}\psi
\Bigg[\frac{3\left|C_{3}\right|^{2}}{-2\delta+4}
\left(1-e^{-(-2\delta+4)N}\right)
\nonumber+\frac{4\delta+5}{
2\delta+6}\left|C_{4}\right|^{2}\left(1-e^{-(2\delta+6)N}\right) \Bigg ] \nonumber\\
\hspace*{-1.25cm}+\frac{24\gamma_{g} H^{4}}{16\pi^{2}}e^{
\pi\delta\gamma_{g}}\psi\left|C_{4}\right|^{2}
\frac{(2\delta+1)\left(\delta+1\right)}{2\delta+5}\left(1-e^{
-(2\delta+5)N}\right)\nonumber\\
\hspace*{-1.25cm}+\frac{2\gamma_{g}\nu H^{4}}{16\pi^{2}}e^{
\pi\delta\gamma_{g}}\psi\frac{
2\delta+1}{2\delta+5}\left|C_{4}
\right|^{2}\left(1-e^{-(-2\delta+4)N}\right),
\end{eqnarray}
where have been used $\psi=\left.\frac{\accentset{\ast}{H}^{2}}{Q}\right|_{\tau = \tau_{f}}$, $\mu=\left.\chi'^{2}\accentset{\ast}{\tau}^2\right|_{\tau = \tau_{f}}$ and $\nu=\left.\chi'^{2}\accentset{\ast}{\tau}\right|_{\tau = \tau_{f}}$.
Notice a slight difference between both frames  in the magnetic and electric density terms, due to the existence of the additional variables   $\psi$ and $\mu$, missing  in the Jordan  frame. This difference lies  in the parameter $\omega(\phi)$ on the Jordan frame  that emerged from  its equivalence with $f(R)$ theories.
The following contour plots display the permitted and forbidden regions for $\rho_{B}$, $\rho_{E}$, and $\rho_{tot}$ for different values of delta, and $\psi$ values running to $0-1$, $\mu$ goes to $0-70$ and $\nu=0.5$. In figure \ref{fig:constraint_1EF},  the permitted values for $\rho_{B}$, and $\rho_{tot}$ are located in regions where $\mu$ is close to zero. The electric density is zero as we can see in the first term  of the equation \ref{eq:densityEF_Example}.
\newline
Figure \ref{fig:Energy_1EF} shows the similarity between $\rho_{B}$ (green line) and $\rho_{tot}$ (blue line) yielding a negligible value of $\Delta\rho$  to avoid increased energy on inflation.  The curve with $\mu=5$ falls rapidly for $\psi=0.2$, because energy densities run into the forbidden region.
\newline
The top  pannel in Figure~\ref{fig:constraint_2EF} shows the behavior of $\rho_{E}$, $\rho_{B}$ while the bottom panel exhibit $\rho_{EB}=\rho_{E}+\rho_{B}$ and $\rho_{tot}$ with $\delta=0$. The forbidden region for $\rho_{B}$  expands quickly  when $\psi$ goes to $0.1$. In this case,  $\Delta\rho$ contributes to constraint $\rho_{tot}$ for the small values of $\psi$.
 Figure \ref{fig:Energy_2EF},  unveils that for $\mu$ greater, the contribution of $\Delta\rho$ affects notably  $\rho_{tot}$. For $\mu=50$ the enhancement  of energy is faster than $\mu=5$. In contrast, in figures \ref{fig:constraint_3EF}-\ref{fig:Energy_3EF} notice the null contribution from $\Delta\rho$ to the total energy density, i.e, $\rho_{tot}$ practically is due to the contribution of $\rho_{E}$. For the latter, the energy grows faster for  $\mu=60$ than $\mu=20$, especially in the range of $\psi$ $0-20$.
\newline

\begin{figure}[!htb]
	\centering % \begin{center}/\end{center} takes some additional vertical space
	\includegraphics[width=\textwidth]{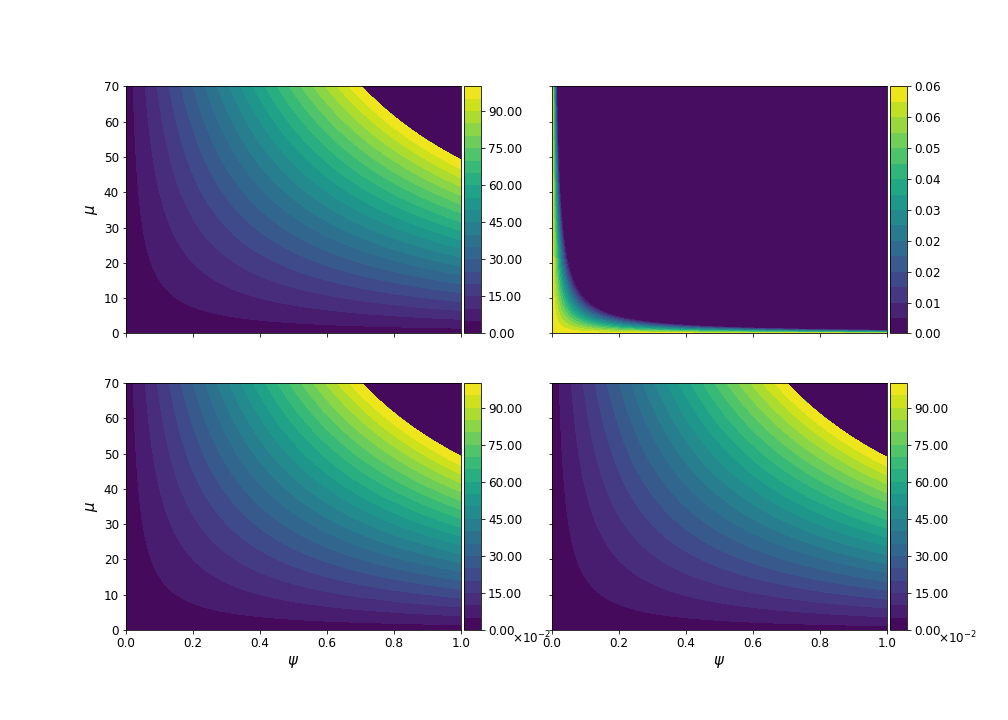}
	% "\includegraphics" is very powerful; the graphicx package is already loaded
	\caption{\label{fig:constraint_4EF} Contour plot with the permitted values and forbidden regions for $\rho_{E}$ (left upper plot), $\rho_{B}$ (right upper plot), $\rho_{EB}$ (left bottom plot) and $\rho_{tot}$ (right bottom plot) with $\delta=2$. Here, the scale is $\times10^{-2}$ and the forbidden values are in purple. We can see two prohibited regions, the first one for small $\mu$ and $\psi$  values, and the second one for $\mu$  above $50$ and $\psi$ higher than $0.6$.}
\end{figure}

\begin{figure}[!htb]
	\centering % \begin{center}/\end{center} takes some additional vertical space
	\includegraphics[width=\textwidth]{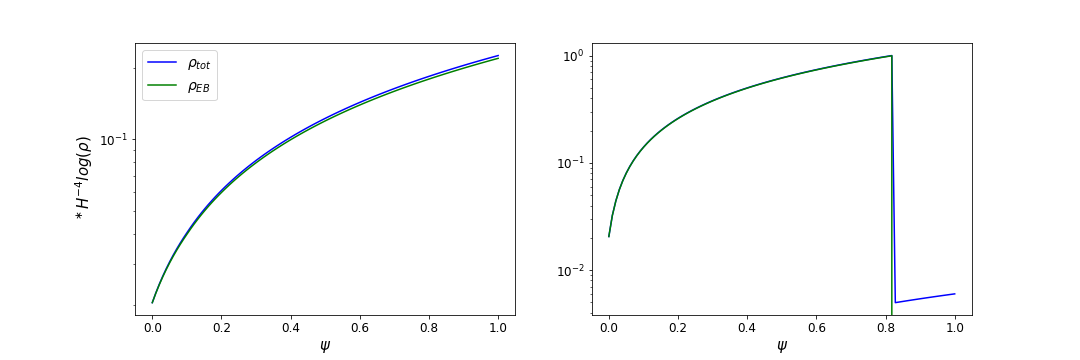}
	% "\includegraphics" is very powerful; the graphicx package is already loaded
	\caption{\label{fig:Energy_4EF}  Plots for $\rho_{EB}$ (green line) and $\rho_{tot}$ (blue line) taking $\mu=10$ (left plot) and  $\mu=60$ (right plot) for $\delta=2$, choosing $\nu=0.5$, $\psi$ ranges $0-1$. }
\end{figure}

\noindent Finally, figures \ref{fig:constraint_4EF}-\ref{fig:Energy_4EF} present the behavior  for  a scale-invariant  magnetic field. Here, $\Delta\rho$ does not contribute to the energy total density.

\noindent Let us now obtain the same observable quantities as we found in the Jordan frame. For the  magnetic spectral index, we have
\begin{equation}\label{eq:nBEF}
\fl \hspace{1cm} 2(2\pi)^{3}\accentset{\ast}{P}_{B}=k^{2}<|{\accentset{\ast}{A}}|^{2}> \Rightarrow k^{\accentset{\ast}{n}_{B}}\propto\accentset{\ast}{P}_{k}\rightarrow \accentset{\ast}{n}_{B}=-2\delta+1 \quad\mbox{for}-k\accentset{\ast}{\tau}<<1,
\end{equation}
using the fact that
\begin{equation}
	\accentset{\ast}{B}(k)=k\accentset{\ast}{A}_{+}(k)=\frac{e^{\pi\gamma_{g}}\sqrt{k}}{2\sqrt{2}\accentset{\ast}{a}\sqrt{Q}}\left(\accentset{\ast}{C}_{3}\left(-k\accentset{\ast}{\tau}\right)^{-\delta}+\accentset{\ast}{C}_{4}\left(-k\accentset{\ast}{\tau}\right)^{\delta+1}\right).
\end{equation}
While  the magnetic field for the present epoch is
\begin{equation}\label{eq:B_actualepoch_EF}
	\accentset{\ast}{B}_{0}=0.63(1-\psi\mu)\times10^{-10}G\left(\frac{{\accentset{\ast}{H}}}{10^{-5}M_{pl}}\right).
\end{equation}
Finally, the helicity in this frame is written as
\begin{equation}
\fl \mathcal{\accentset{\ast}{H}}=\int \accentset{\ast}{A}\cdot \accentset{\ast}{B} d^{3}x\Rightarrow\mathcal{\accentset{\ast}{H}}=\frac{e^{\delta\gamma_{g}\pi}}{16\pi^{2}\accentset{\ast}{a}^{2}Q}\left[\frac{|C_{3}|^{2}}{-2\delta+3}(-k\accentset{\ast}{\tau})^{-2\delta}k^{3}+\frac{|C_{4}|^{2}}{2\delta+5}(-k\accentset{\ast}{\tau})^{2\delta+2}k^{3}\right],
\end{equation}
while the kinetic helicity spectral index reads as
\begin{equation}
	\accentset{\ast}{n}_{H}=-2\delta+2.
\end{equation}
As we can see from  equations (\ref{eq:B_actualepoch_EF}) and (\ref{eq:B_actualepoch_JF}), a disparity between both frames is clear.

\section{Discussion about Jordan and Einstein frames}\label{discussionFrames}
% In this section, we want to compare the amplitude of the magnetic field in Jordan and Einstein frames.
Through this paper, we have shown differences in distinct quantities in scalar-tensor theories between  Jordan and Einstein frames in the context of primordial magnetic fields, by using a model of power law coupled to $F^{2}$ and $F\tilde{F}$. We found out that the magnetic spectral index and its helicity  are similar in both frames. In contrast, the amplitude of the magnetic field today differs on each frame (see \ref{eq:ratiosB}). It is important to remark that different assumptions have been taken along the work to find the above results, limiting the solutions that we have encountered. The primary goal of this paper was to study the equivalence between Jordan and Einstein frames  with the purpose of comparing observables between them via  asymptotic solutions found in the scenario of magnetogenesis. For more detail about the evolution of the electromagnetic field during inflation, see \cite{Savchenko_2018, Durrer_2022}.  \newline
Finally, the ratio between the amplitudes of the magnetic field in Jordan and Einstein frames in the actual epoch (assuming a scale-invariant case) is written as
%where we have used the equations (\ref{eq:B_actualepoch_JF}) and (\ref{eq:B_actualepoch_EF}),
\begin{equation}\label{eq:ratiosB}
 \frac{\accentset{\ast}{B}_{0}}{B_{0}}=(1-\psi\mu)\frac{\accentset{\ast}{H}}{H}.
\end{equation}

\begin{figure}
	\centering % \begin{center}/\end{center} takes some additional vertical space
	\includegraphics[width=0.5\textwidth]{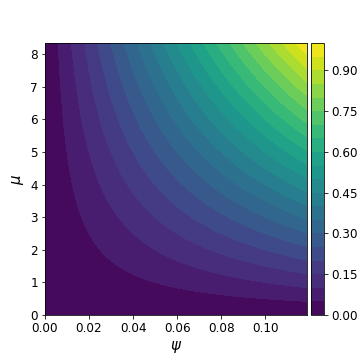}
	% "\includegraphics" is very powerful; the graphicx package is already loaded
	\caption{\label{fig:mupsi}  Ratio between Hubble parameters in JF to EF using the same amplitude of the magnetic field taking scale-invariant case. $\mu$ goes to $0-8$ and $\psi$ runs over $0-0.12$.
	}
\end{figure}
\noindent This equation depends on two factors, the constraints values (see figure \ref{fig:constraint_4EF}), and the Hubble parameters in both frames. To show an approximate relation between the Hubble parameters in each frame, let us take two values $\mu$ and $\psi$, from figure \ref{fig:mupsi} (this figure is a zoom of the permitted values for $\mu$ and $\psi$ of the $\rho_{B}$ values). For example, assuming $\mu=5$ and $\psi=0.1$, and a value of $B_{0}=10$ nG, we can find that, $\accentset{\ast}{H} = H$ with $\accentset{\ast}{B}_{0}=5$ nG.

\begin{figure}
	\centering % \begin{center}/\end{center} takes some additional vertical space
	\includegraphics[width=0.8\textwidth]{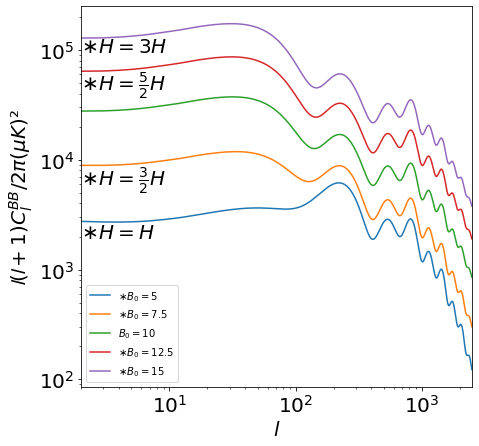}
	% "\includegraphics" is very powerful; the graphicx package is already loaded
	\caption{\label{fig:B_changing_nb_fix}  The B-mode spectrum from the PMF vector mode, $B_{0}=10 nG$in JF, and different values of the magnetic field in EF $\ast^{}B_{0}=5,7.5,12.5,15$ nG. }
\end{figure}
\noindent In the figure \ref{fig:B_changing_nb_fix} (this figure was realized using a patch of the CAMB code account for primordial magnetic field \cite{Zucca_2017, Lewis_2000}), we can see the relation between these two parameters for one value of the amplitude of the magnetic field in JF, $B_{0}=10$ nG, and different values of the magnetic field in EF, $\accentset{\ast}{B}_{0}=5,7.5,12.5,15$ nG, taking $n_{b}=-2.9$.

\section{Conclusions}\label{conclusions}
 It still remains an open question about  the equivalence between Jordan and Einstein frames, and whether there exists a  physically privileged frame. This paper has addressed the problem of performing calculations in both frames  under the primordial magnetic field cosmological scenario.  We have calculated the electromagnetic energy density in both frames, where the electric and  magnetic energy densities along with other contributions  from   couplings between the gravity sector with the electromagnetic field tensor contribute to the total energy density. Assuming a power law model in the magnetic spectra, we found that in Jordan frame the electric and magnetic energy densities only depend on the power $\xi$.  In contrast,  the total energy density in the Einstein frame  depends not only on the power of the coupling  but also, on additional  parameters relevant  to not spoiling inflation energy (\ref{eq:densityEF_Example}). The amount $\Delta\rho$ (the other contributions of the energy density) was restricted in both frames. For instance,  in the Jordan frame, the parameters that we found were $\alpha$ and $\beta$, while  in the Einstein frame, we used  $\psi$ and $\nu$. The $\mu$ and $\nu$ terms result because when we do the equivalence between scalar-tensor theory in Jordan frame with $f(R)$-gravity we turn off the $\omega$ term but this does not happen in the other frame.  We obtained the same value for which the magnetic field is scale-invariant, and we derived a relation to the present magnetic field in both frames in the case of scale-invariant (\ref{eq:ratiosB}).  We expect that some of the results presented in the paper contribute to the ongoing discussion on the relationship between these two frames.

\section{Acknowledgments}
Joel Velásquez and Leonardo Castañeda were supported by Patrimonio Autónomo - Fondo Nacional de Financiamiento para la Ciencia, la Tecnología y la Innovación Francisco José de Caldas 
(MINCIENCIAS - COLOMBIA) Grant No. 110685269447 RC-80740-465-2020, projects 69723
\section*{References}
\bibliography{references}

\end{document}